# The earth as a living planet: Human-type diseases in the earthquake preparation process.

Y. F. Contoyiannis[a], S. M. Potirakis[b], and K. Eftaxias[a]

*a. Department of Physics, Section of Solid State Physics, University of Athens, Panepistimiopolis, GR-15784, Zografos, Athens, Greece, ceftax@phys.uoa.gr .*
*b. Department of Electronics Engineering, Technological Education Institute (TEI) of Piraeus, 250 Thivon & P. Ralli, GR-12244, Aigaleo, Athens, Greece, spoti@teipir.gr .*

**Abstract**

The new field of complex systems supports the view that a number of systems arising from disciplines as diverse as physics, biology, engineering, and economics may have certain quantitative features that are intriguingly similar. The earth is a living planet where many complex systems run perfectly without stopping at all. The earthquake generation is a fundamental sign that the earth is a living planet. Recently, analyses have shown that human-brain-type disease appears during the earthquake generation process. Herein, we show that human-heart-type disease appears during the earthquake preparation of the earthquake process. The investigation is mainly attempted by means of critical phenomena, which have been proposed as the likely paradigm to explain the origins of both heart electric fluctuations and fracture induced electromagnetic fluctuations. We show that a time window of the damage evolution within the heterogeneous Earth's crust and the healthy heart's electrical action present the characteristic features of the critical point of a thermal second order phase transition. A dramatic breakdown of critical characteristics appears in the tail of the fracture process of heterogeneous system and the injury heart's electrical action. Analyses by means of Hurst exponent and wavelet decomposition further support the hypothesis that a dynamical analogy exists between the geological and biological systems under study.

## 1. Introduction

Empirical evidence has been accumulated supporting the view that a number of systems arising from disciplines as diverse as physics, biology, engineering, and economics may have certain quantitative features that are intriguingly similar (Bar-Yam, 1997). Characteristically, de Arcangelis et al. (2006) presented evidence for universality in solar flare and earthquake (EQ) occurrence. Kossobokov and Keilis-Borok (2000) have explored similarities of multiple fracturing on a neutron star and on the Earth, including power-law energy distributions, clustering, and the symptoms of transition to a major rupture. Sornette and Helmstetter (2002) have presented occurrence of finite-time singularities in epidemic models of rupture, EQs, and starquakes. Abe and Suzuki (2004) have shown that internet shares with EQs common scale-invariant features in its temporal behaviors. Peters et al. (2002) have shown that the rain events are analogous to a variety of non-equilibrium relaxation processes in Nature such as EQs and avalanches. Fukuda et al. (2003) have shown similarities between communication dynamics in the Internet and the automatic nervous system.

Importantly, strong analogies between the dynamics of geophysical and biological systems have been reported by numerous authors. Indeed, authors have proposed that dynamics of EQs and neurodynamics could be analyzed within similar mathematical frameworks (Herz and Hopfield, 1995; Rundle et al., 2002). Characteristically, driven systems of interconnected blocks with stick-slip friction capture the main features of EQ process. These models, in addition to simulating the aspects of EQs and frictional sliding, may also represent the dynamics of neurological networks (Herz and Hopfield, 1995, and references therein). Hopfield (1994) proposed a model for a network of $N$ integrate-and-fire neurons. In this model, the dynamical equation of $k^{th}$ neuron, see equation (28) in (Hopfield, 1994) is based on the Hodgekin-Huxley model for neurodynamics and represents the same kind of mean field limit that has been examined in connection with EQs (Rundle et al., 2002). Recently, Osorio et al. (2010) in a pioneering work under the title "Epileptic seizures: Quakes of the brain?" have shown that a dynamical analogy supported by scale-free statistics exists between epileptic seizures and EQs. More precisely, the authors performed the analysis using: (i) a population of different EQs, and (ii) a population of different epileptic seizures (ESs). In recent works we have shown by means of nonextensive statistical mechanics the existence of a dynamical analogy between epileptic seizures and EQs at the level of a single fault / seizure activation, namely, we have shown that a dynamical analogy exists for the ways in which firing neurons / opening cracks organize themselves to produce *a single* epileptic seizure / EQ (Li et al., 2005; Kapiris et al., 2005; Eftaxias et al., 2006, 2012). This dynamical analogy by means of nonextensivity is extended to earthquakes, magnetic storms and solar flares (Balasis et al., 2011a, 2011b).

It has been suggested that the underlying mechanism behind human heart rate regulation shares the general principles of other complex systems (Lin and Hughson, 2001 and references therein). Picoli et. al. (2007) reported similarities between the dynamics of geomagnetic signals and heartbeat intervals. Herein, we investigate the existence of similarities between a geophysical pre-seismic signal and a biological signal. We focus on the dynamics of: (i) fracture induced electromagnetic (EM) radiations rooted in two well established distinct stages of fracture of heterogeneous medium in the field, and (ii) human electrocardiogram (ECG) rooted in healthy and injury heart's electrical action.

Several techniques have been applied to detect high risk patients from ECG (Varotsos et al., 2007 and references therein). A method which not only identifies the risk but also provides an estimate of the time of an impending cardiac arrest has been recently proposed based on the analysis of electrocardiograms by means natural time (Varotsos et al., 2007, 2011a, 2011b). Herein, the investigation is mainly attempted by means of critical phenomena. Importantly, critical phenomena have been proposed as the likely paradigm to explain the origins of both heart rate fluctuations and fracture induced EM fluctuations, suggesting that the theory of phase transitions and critical phenomena may be useful in elucidating the mechanism of their complex dynamics (Kiyono et al., 2004, 2005; Contoyiannis et al., 2004, 2005). One possible reason for the reported similarities concerns the way in which correlations spread thought a system comprised of subunits. Nature seems to paint the following picture: first, single isolated activated parts emerge in the system which, subsequently, grow and multiply. This leads to cooperative effects. Long-range correlations build up through local interactions until they extend throughout the entire system. *The challenge is to determine the "critical epoch" during which the "short-range" correlations evolve into "long-range" ones.* Therefore, the theory of phase transitions and critical phenomena seem to be useful in elucidating similarities in different complex systems.

Characteristic features at a critical point of a second order transition are the existence of strongly correlated fluctuations and the scale invariance in the statistical properties. More precisely, experiments and also calculations on mathematical models confirm that right at the "critical point" the subunits are well correlated even at arbitrarily large separation, namely, the probability that a subunit is well correlated with a subunit at distance $r$ away is unity and the correlation function $C(r)$ follows long-range power-law decay. At the critical state appear self-similar structures both in time and space. This fact is mathematically expressed through power law expressions for the distributions of spatial or temporal quantities associated with the aforementioned self-similar structures (Stanley, 1987, 1999).

Herein, analyzing human ECG time-series of healthy individuals, we provide evidence that the heart dynamics include the aforementioned critical features. Moreover, we show that beyond the appearance of a broad range of integrated self-affine outputs and long-range correlations, the fluctuations of the healthy ECG time-series are also characterized by strong anti-persistence behavior, namely, an underlying negative feedback mechanism that kicks the system far from extreme operation. The aforementioned crucial features characterize a healthy system, since such a mechanism provides *adaptability*, the ability to respond to various stresses and stimuli of everyday challenges (Goldberger et al., 2002). Turning focus on another complex system, fracture induced MHz EM fields permit the real-time monitoring of the gradual damage of stressed heterogeneous materials in the Earth's crust as an earthquake is approaching, as it happens at the laboratory fracture experiments (Contoyiannis et al., 2005; Lacidogna et al, 2011). We have shown that at some time point of the damage evolution within the Earth's crust a critical time window emerges which appears all the aforementioned features of critical point of a thermal second order phase transition. During this epoch the stressed crust is in a "healthy state" for the reason that it includes critical characteristics, thus keeping the system away from extreme states.

On the other hand, below and above of the critical point of a continuous thermal phase transition a dramatic breakdown of critical characteristics, in particular long-range correlations, appears; the correlation function turns into a rapid exponential decay (Stanley, 1987, 1999). We show that "injury" states of heart include the characteristic features of the state which is away from the critical point. Moreover, the system shows lower anti-persistent behavior. Thus, the system has lost a part of its adaptability, namely, the ability to respond to all stresses and stimuli. Importantly, a time window is launched in fracture induced EM time series after the critical window appearance which has exactly the same "injury" characteristics. The fracture has been restricted in a narrow zone completing the fracture of heterogeneous medium. We note that our finding can be supported by other studies comparing the heart dynamics of healthy people and congestive heart failure patients as well as by the heart dynamics of healthy frogs (Contoyiannis et al., 2004). The analysis of time-series has been performed in terms of the recently introduced method of critical fluctuations (MCF) of thermal systems that undergo a second order phase transition (Contoyiannis et al., 2002, 2004, 2005, 2007, 2008, 2010), Hurst exponent (Hurst, 1951; Mandelbrot and Wallis, 1968), wavelet scalogram (e.g., Pinsky, 2002; Stark, 2005; Weeks, 2007), and wavelet marginal spectrum (e.g., Yeh and Liu, 2008; Di Marco et al., 2012).

The remaining of this contribution is organized as follows. In Sec. 2 we provide concisely the necessary background knowledge on the MCF, Hurst exponent, wavelet scalogram, and wavelet marginal spectrum. In Sec. 3 we proceed to the analysis of normal vs. pathological subjects by means of MCF for both heart time-series and pre-seismic EM emissions time-series. In Sec. 4 we proceed to the analysis of normal vs. pathological subjects by means of Hurst exponent and wavelet decomposition. Finally, in Sec. 5 we summarize and discuss our results.

## 2. Analysis methods

In the following we present in brief the key points of the analysis methods employed in this contribution. More specifically, the method of critical fluctuations (MCF) is introduced, while the notions of Hurst exponent, $H$, and energy density of the wavelet marginal spectrum, $P_{d,\Delta T}(a)$, are shortly defined.

### 2.1 The method of critical fluctuations

Recently we have introduced the method of critical fluctuations (MCF), as a method which can reveal the critical state as well as the departure from critical state. The MCF is described in detail in (Contoyiannis et al., 2002). In a few words, for a given stationary time-series this method estimates the distribution of waiting times $l$ in a properly defined laminar region. It is known (Schuster, 1998) that this distribution should follow power-law decay for a system possessing intermittent dynamics. On the other hand, the corresponding waiting times distribution of a conventional (noncritical dynamics) real data time-series follows an exponential decay, rather than a power-law one (Contoyiannis et al., 2004), due to stochastic noise and finite size effects.

Therefore, a function which combines both power-law and exponential decay is more appropriate to fit the distribution of waiting times in a log-log scale diagram. Such a function is of the form of $\rho(x)$:

$$\rho(x) = Cx^{-p_2}e^{-p_3 x} \tag{1}$$

As a result, one can estimate the dynamics of a real time-series by fitting the waiting times distribution using $\rho(x)$. The estimated values of the two characteristic exponents $p_2$ and $p_3$, reveal the underlying dynamics.

The first step of the analysis is the acquisition of the time-series values histogram, on which we search, following the method of turning points (Pingel et al., 1999; Diakonos et al., 1999), for the marginal unstable fixed-point, which it is considered as the begin of laminar regions. According to this method, the fixed point lies in the vicinity of the abrupt variation edge of the histogram.

In a second step, we are searching for a plateau (or a region with slow variation) in the histogram. This is defined as the laminar region, which decays slowly. The end of the laminar region is treated as a free parameter determining the size of laminar region. The waiting times in the laminar region, are characterized as laminar lengths $l$. We successively calculate laminar lengths and then obtain their distribution.

Finally, a fitting of the obtained distribution is performed on a log-log plot ending-up to the estimation of the $p_2$ and $p_3$ exponents. The application of the MCF method is presented step-by-step in the examples analyzed in SubSec. 3.1.

The above described three-step procedure is repeated for other ends of the laminar region and finally the evolution of the $p_2$ and $p_3$ exponents vs. the end of laminar region is plotted. According to the discussion in (Contoyiannis et al., 2007) only values of exponent $p_2$ greater than 1 and values of exponent $p_3$ close to zero indicate proximity to critical behavior in terms of a second order phase transition. If $p_2 < 1$ then usually $p_3$ is increased and a departure from criticality is signaled.

So far, the MCF has been applied on numerical experiments of thermal systems (Ising models) (Contoyiannis et al., 2002), on electromagnetic pre-seismic signals (Contoyiannis et al., 2005, 2008, 2010), and on electrocardiac signals from biological tissues (Contoyiannis et al., 2004).

Recently, we have shown that the laminar length distribution produced by a time-series from ECGs of healthy frogs demonstrated power-law behavior like the corresponding distributions produced by the critical fluctuations of thermal system on the critical point of a second order phase transition (Contoyiannis et al., 2004). We have interpreted this result as a manifestation that the healthy heart of the frog behaves as a physical system in its critical point. In other words, there is time scale invariance, namely the healthy heart can respond to all the temporal scale excitations in the same way. Almost simultaneously, another team reported that the healthy human heart demonstrates characteristic features at a critical point of a second order phase transition (Kiyono et al., 2005), a result similar to ours. We note that this team analyzed heartbeat time-series.

## 2.2 Hurst exponent

Roughness of a geometry or a time-series, or long-term memory, in a complex process is usually expressed in terms of the well known Hurst exponent, initially introduced for the analysis of hydrological data (Hurst, 1951; Mandelbrot and Wallis, 1968) in the form of rescale range ($R/S$) analysis.

The $R/S$ analysis is based on two quantities: first, the range $R_n$, which is the difference between the maximum and minimum values of the accumulated departure of the time series from the mean, calculated over each one ($n = 1, 2, ..., d$) of the $m-$samples long sub-series in which the time-series can be divided, and second, the standard deviation of the corresponding sub-series $S_n$. The so-called rescaled range is exactly the ratio of $R$ by $S$. Hurst found that ($R/S$) scales by power - law as time (i.e., the sample length $m$ of the sub-series) increases,

$$(R/S)_m \propto m^H, \qquad (2)$$

where $H$ is the Hurst exponent, an empirical relation well describing a variety of time series of natural phenomena. The exponent $H$ is estimated as the linear slope of a $\log(R/S)_m - \log m$ representation.

## 2.3 Wavelet scalogram and wavelet marginal spectrum

The frequency analysis of non-stationary signals has been initially addressed by the introduction of the Short-Time Fourier Transform (STFT) and the associated time-frequency representation which is known as a "spectrogram" (Nawab and Quatieri, 1988; Oppenheim et al., 1999). An advance on STFT is considered to be the continuous wavelet transform (CWT) (Pinsky, 2002). A detailed comparison of the STFT and the WT can be found in (Kim and Kim, 2001). The Wavelet transform (WT) is a kind of multi-resolution analysis (analysis of the signal at different frequencies with different resolutions) and is defined as the convolution of the signal under analysis with a set of wavelet functions generated by the "mother wavelet" by dilation and translation (Daubechies, 1992; Mallat, 1998; Stark, 2005):

$$W_\psi^x(a,b) = \frac{1}{\sqrt{c_\psi |a|}} \int_{-\infty}^{+\infty} x(t) \cdot \psi^* \left( \frac{t-b}{a} \right) dt, \quad a \neq 0, t \in \Re. \qquad (3)$$

$\psi(t)$ is the mother wavelet function, while $\psi_{a,b}(t) = \frac{1}{\sqrt{|a|}} \psi\left(\frac{t-b}{a}\right)$ is the wavelet, i.e., the dilated and translated version of the mother wavelet, $a$ is called the scale or the dilatation parameter, $b$ is called translation or offset parameter or shift or lag, and $\frac{1}{\sqrt{c_\psi |a|}}$ is a normalization factor, which ensures the satisfaction of the admissibility condition and the energy normalization throughout scales (Stark, 2005). Several options have been proposed for the calculation of the discretized CWT, e.g., (Torrens and Compo, 1998; Stark, 2005; Rioul and Duhamel, 1992; Weeks, 2007).

The wavelet analogue to the spectrogram is the *scalogram*, defined as:
$$Scalogram(x) = \left|W_\psi^x(a,b)\right|^2. \tag{4}$$

Therefore, the resulting visual representation of a scalogram is a function of scale ($a$) and time ($b$). Usually employed mother wavelets are the Haar, the Morlet, the Mexican Hat and the Meyer functions (Daubechies, 1992).

The wavelet scale has similar meaning to the scale in maps, i.e., large scale corresponds to "overall view", or long term behavior, while small scale corresponds to "detail view", or local behavior. It could be considered corresponding to the inverse to "frequency" if one likes to refer to Fourier analysis terms. WT has proved to be an ideal tool for measuring singularities based on the decay/growth rate of wavelet transform coefficients with respect to scales (Mallat and Hwang, 1992; Li and Liner, 2005).

The scale marginal $P_{\Delta T}(a)$ of the power in the time interval $\Delta T = [t_1, t_2]$ (or marginal energy density with respect to scale over a time interval) is frequently called "the wavelet marginal spectrum" and is obtained by integration over time interval $\Delta T$ (Di Marco et al., 2012; Yeh and Liu, 2008)
$$P_{\Delta T}(a) = \sum_{b=t_1}^{t_2} \left|W_\psi^x(a,b)\right|^2. \tag{5}$$

We investigate here the energy density of the wavelet marginal spectrum as a discrete probability distribution, $P_{d,\Delta T}(a)$,
$$P_{d,\Delta T}(a) = \frac{P_{\Delta T}(a)}{\sum_a P_{\Delta T}(a)}. \tag{6}$$

### 3. Normal vs. pathological subjects by means of MCF

In this section we study the existence of similarities in the breakdown that comes with the transition from the critical healthy epoch to the pathological one in both the biological and the geophysical signals under study. The analysis is performed by means of the MCF method.

### 3.1 Focus on the human heart dynamics.

In this sub-section we investigate the application of the MCF method on ECGs recorded on human subjects both healthy and suffering from heart infraction. The data (ECGs) were taken from the Physikalisch-Technische Bundesanstalt (PTB) Diagnostic ECG Database[1] (Bousseljot et al., 1995; Kreiseler and Bousseljot, 1995; Goldberger et al., 2000).

---

[1] http://physionet.ph.biu.ac.il/physiobank/database/ptbdb/

In the following we present the application of the MCF method on a healthy human ECG labeled as "p121" and a representative example of human heart myocardic infraction under the code name "p023". The step-by-step procedure is described in detail as follows:

- First, the original ECG data were pre-processed in order to ensure a minimum cumulative stationarity of the time-series to be analyzed. The times-series under analysis, $V(t)$, were obtained by restricting to the lower fluctuations, i.e., clipping the high peaks of the original time-series. The ECG of "p121" (healthy) and "p023" (infraction) are shown in Figs 1a and 1e, respectively. The analyzed times-series correspond to the recorded voltage values which fall within the interval $[-0.08, 0.3]$ and $[-0.2, 0.3]$, correspondingly.

- A typical characteristic of the histogram of a time-series obeying to critical intermittent dynamics is the existence of a plateau (or slow variation) region which decays, namely laminar region. In Figs 1b and 1f the voltage-values distributions (histogram) of the analyzed time-series are shown. The above mentioned plateau region which usually appears in MCF is clearly seen.

- The fixed point according to the method of turning points lies in the vicinity of the abrupt edge of the distribution (Pingel et al., 1999; Diakonos et al., 1999). From the distributions of Figs 1b and 1f the corresponding fixed points are identified to be $V = -0.06$ and $V = -0.15$, respectively. The end point, $V_l$, of the laminar region is treated as a free parameter, usually, inside the decay region.

- Then, the laminar lengths, $l$, are produced as the lengths that result from successive $V$-values obeying the condition $V_o \leq V \leq V_l$. In others words, the laminar lengths are the waiting times inside the laminar region.

- Finally, the resulting laminar distributions $P(l)$ are plotted in log-log scale and from there the exponents $p_2$, $p_3$ are estimated by fitting the function of Eq. (1). The plots of the exponents $p_2$, $p_3$ vs $V_l$ (the end of laminar regions) are plotted in Figs 1c and 1g for the healthy and the infraction case, in that order. A representative example of the corresponding laminar distributions for each one of the examined cases is given in Fig 1d and Fig. 1h, considering $V_l = -0.023$ for the healthy case and $V_l = -0.1$ for the infraction case. The continuous lines in these plots represent the fitted functions of the form of Eq. (1). The estimated values of the involved exponents are $p_2 = 1.32$ and $p_3 = 0.03$ for the healthy case, while $p_2 = 0.5$ and $p_3 = 0.13$ for the pathological case.

As it can be seen from Fig. 1c all the estimated values for the exponents $p_2$ and $p_3$ obey the condition of criticality ( $p_2 > 1, p_3 \cong 0$ ) in the healthy case. These results lead us to conclude that the healthy heart is very close to a critical state in terms of second order phase transition. Note that this is a result similar to the healthy frog's heart result (Contoyiannis et al., 2004). On the other hand, the $p_2$, $p_3$ exponents values shown in Fig. 1g signify that the state of the pathological case is far from critical state. The timescale invariance of the heart tissue has been lost. This means that the heart of the myocardic infraction subject is not able to respond to the excitations of all time-scales.

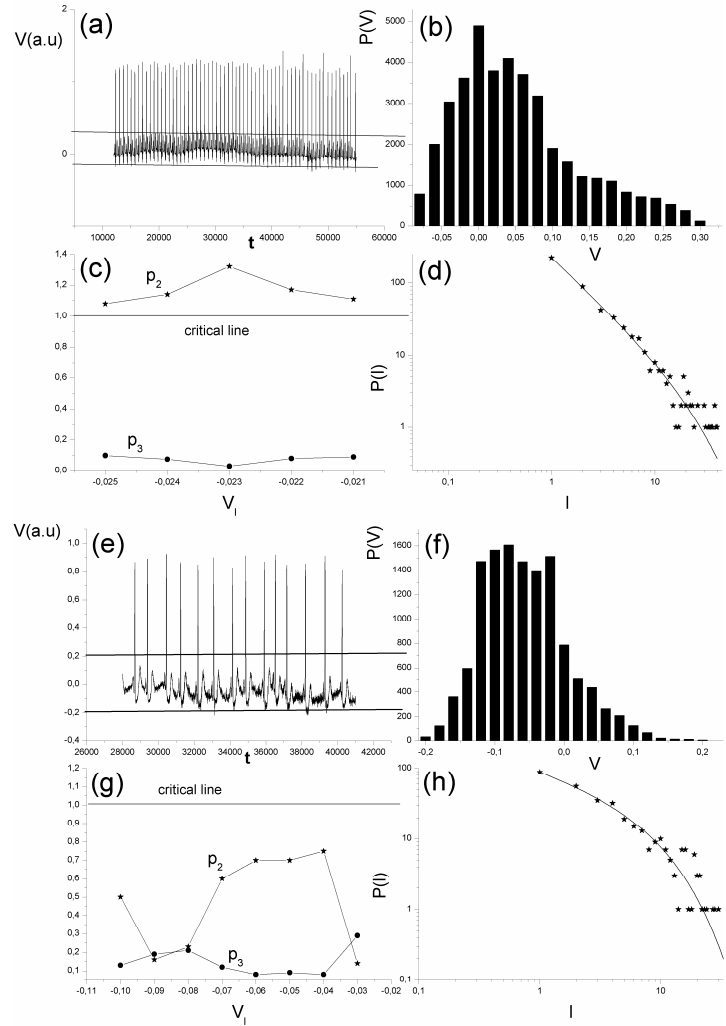

**Fig. 1**. (a) The ECG of the healthy subject "p121". The analyzed times-series corresponds to the recorded voltage values which fall within the interval $[-0.08, 0.3]$. (b) The voltage-value distribution of the time-series of Fig. 1a. From this distribution, the fixed point has been determined as $V = -0.06$. (c) The "healthy" exponents $p_2$, $p_3$ vs. $V_l$ (the end of laminar regions). (d) A representative example of a "healthy" laminar distribution for $V_l = -0.023$. (e) The ECG for a life-threatening condition of the subject "p023". The analyzed times-series corresponds to the recorded voltage values which fall within the interval $[-0.2, 0.3]$ (f) the voltage-values distribution of the time-series of Fig. 1e. From this distribution, the fixed point has been determined as $V = -0.15$. (g) The "pathological" exponents $p_2$, $p_3$ vs. $V_l$ are shown. (h) A representative example of a "pathological" laminar distribution for $V_l = -0.1$.

### 3.2 Focus on earthquake dynamics through pre-seismic EM data

During an EQ preparation process EM emissions in MHz and kHz band appear one after the other a few days or a few hours prior to the occurrence of the EQ event, exactly as it happens in the laboratory (Eftaxias et al., 2001, 2011; Kapiris et al., 2004; Contoyiannis et al., 2004, 2005; Karamanos et al., 2006; Papadimitriou et al., 2008; Potirakis et al., 2012a, 2012b, 2012c). *The first emerging MHz EM activity is organized during the fracture of heterogeneous medium surrounding the backbone of large and strong asperities distributed along the fault sustaining the system (Contoyiannis et al., 2005).* The finally emerging kHz EM anomaly is thought to be due to the fracture of family of the asperities themselves (Eftaxias et al., 2001; Contoyiannis et al., 2004, 2005; Potirakis et al., 2012b; Minadakis et al., 2012).

The MCF has been applied in a number of cases on pre-seismic EM emissions of the MHz band (Contoyiannis et al., 2005, 2008, 2010). The corresponding MHz EM recordings present intervals on which the application of MCF reveals a critical behavior. These time intervals are called *critical windows* (CWs) and their organization is accomplished in terms of the critical point of a second order phase transition (Contoyiannis et al., 2005).

In the following, the sequence of critical-noncritical windows in MHz EM time series is presented through two examples which refer to two strong EQ events.

The first example concerns the Kozani-Grevena EQ, occurred in 13 May 1995 in North Greece with a magnitude of 6.5. A critical window, recorded 11 hours before the occurrence of the EQ event, is shown in Fig. 2a. This window has duration of about 6.4 h. The recorded amplitude ($\phi$- values) distribution of the CW is shown in Fig. 2b. From this distribution a fixed point $\phi_o$ of about 385mV results. Following the MCF procedure the exponents $p_2$, $p_3$, were estimated for various exit points $\phi_l$. The exponents $p_2$, $p_3$ vs. $\phi_l$ are shown in Fig. 2c. As it can be observed, all the exponents' values obey to critical condition ($p_2 > 1, p_3 \cong 0$) in this "healthy" case. Therefore, the time-series of Fig. 2a is indeed a critical window. A representative example of the Eq (1) fitting for $\phi_l = 220$ is shown in Fig. 2d. The continuous line corresponds to the fitted function of the form of Eq. (1). The estimated values of the involved exponents in this example are $p_2 = 1.21$ and $p_3 = 0.02$.

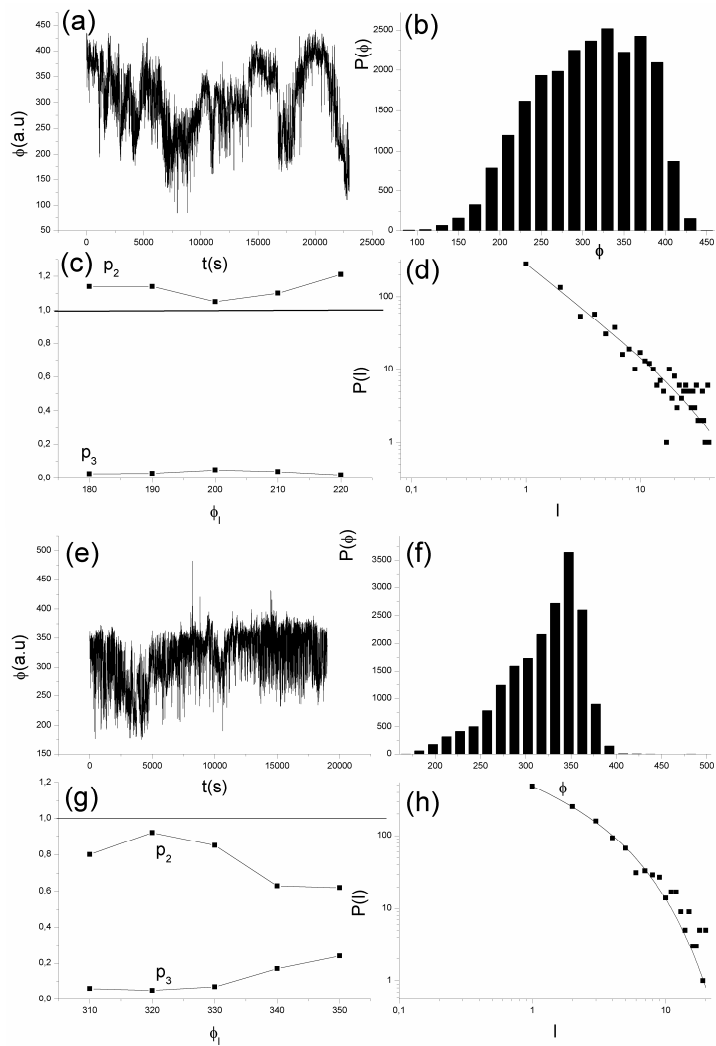

**Fig. 2.** (a) The CW almost 6.4h long associated with the Kozani-Grevena EQ (b) The histogram of the $\phi$-values of Fig. 2a. (c) The corresponding "healthy" exponents $p_2$, $p_3$ vs. the end of laminar region $\phi_l$. The critical limit ($p_2 = 1$) is also depicted. (d) A representative example of a laminar distribution of the CW for $\phi_l = 220$. (e) A noncritical excerpt of the recorded EM time-series, following short after the CW of Fig. 2a, almost 5.5h long (f) The histogram of $\phi$-values of Fig. 2e. (g) The "pathological" exponents $p_2$, $p_3$ vs. the end of laminar region $\phi_l$. The critical limit ($p_2 = 1$) is also depicted. (h) A representative example of a laminar distribution of the noncritical excerpt for $\phi_l = 350$. The departure from critical state is clear.

Short after the CW, ~2.5h later, a clearly noncritical window appeared. The specific noncritical excerpt of the recorded EM time-series, following short after the CW of Fig. 2a, almost 5.5h long, is presented in Fig. 2e. The distribution of $\phi$-values is demonstrated in Fig. 2f. The corresponding fixed point is identified to be $\phi_o = 380$. Fig. 2g ($p_2$, $p_3$ vs. $\phi_l$) indicates that the specific time-series corresponds to a state far from critical, a "pathological" heart analogous. A representative example of MCF analysis is shown in Fig. 2h. In this example, the associated exponents were estimated to be $p_2 = 0.62$ and $p_3 = 0.24$ for the exit point $\phi_l = 350$. A few hours after the specific sequence of critical-noncritical window the strong ($M = 6.5$) EQ event took place.

The second example refers to the L'Aquila EQ, occurred in Central Italy on 6 April 2009 ($M = 6.3$). MHz EM emissions recorded prior to the specific EQ have already been presented (Contoyiannis et al., 2010), however different excerpts were analyzed. Herein, a sequence of critical-noncritical window recorded almost 3 days before the EQ event is presented for the first time.

A critical window of the MHz recordings associated to the L'Aquila EQ is shown in Fig. 3a. This window has duration of about 2.7 h. The corresponding amplitude ($\phi$- values) distribution of the CW is shown in Fig. 3b. Due to the fact that this distribution is symmetric one can select whatever edge as the fixed point. Following the MCF procedure the exponents $p_2$ and $p_3$ values were estimated, for various exit points $\phi_l$; the exponents $p_2$, $p_3$ values vs. $\phi_l$ are shown in Fig. 3c. It is observed that all the exponents' values obey the critical condition ($p_2 > 1$, $p_3 \cong 0$). Therefore, the time-series of Fig. 3a could be characterized as a critical window. A representative example of the Eq (1) fitting for $\phi_l = 780$ is shown in Fig. 3d. The continuous line corresponds to the fitted function of the form of Eq. (1). The estimated values of the involved exponents in this example are $p_2 = 1.18$ and $p_3 = 0.026$.

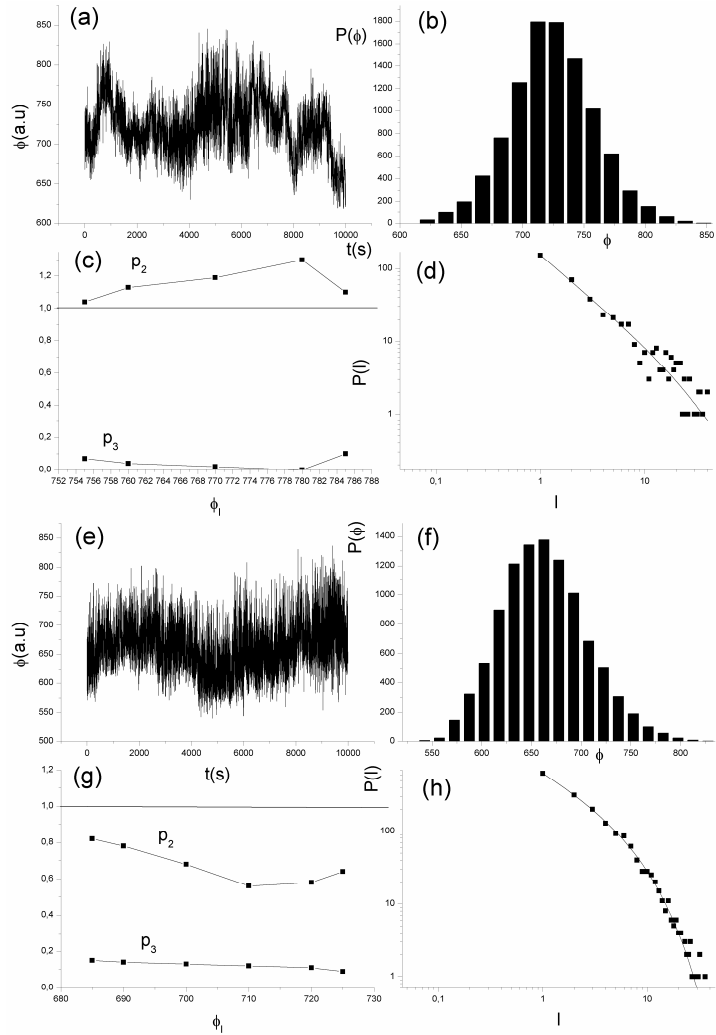

**Fig. 3.** (a) The CW almost 2.7h long associated with the L'Aquila EQ (b) The histogram of the $\phi$-values of Fig. 3a. (c) The corresponding "healthy" exponents $p_2$, $p_3$ vs. the end of laminar region $\phi_l$. The critical limit ($p_2 = 1$) is also depicted. (d) A representative example of a laminar distribution of the CW for $\phi_l = 780$. (e) A noncritical excerpt of the recorded EM time-series, following short after the CW of Fig. 3a, almost 2.7h long (f) The histogram of $\phi$-values of Fig. 3e. (g) The "pathological" exponents $p_2$, $p_3$ vs. the end of laminar region $\phi_l$. The critical limit ($p_2 = 1$) is also depicted. (h) A representative example of a laminar distribution of the noncritical excerpt for $\phi_l = 690$. The departure from critical state is clear.

Short after the CW, ~2.7h later, a clearly noncritical window appeared. The specific noncritical excerpt of the recorded EM time-series, following short after the CW of Fig. 3a, almost 2.7h long, is presented in Fig. 3e. The distribution of $\phi$-values is demonstrated in Fig. 3f. The corresponding fixed point is identified to be $\phi_o = 600$. Fig. 3g ($p_2$, $p_3$ vs. $\phi_l$) indicates that the specific time-series corresponds to a state far from critical, a "pathological" heart analogous. A representative example of MCF analysis is shown in Fig. 3h. In this example, the associated exponents were estimated to be $p_2 = 0.78$ and $p_3 = 0.14$ for the exit point $\phi_l = 690$.

We observe that the MCF reveals that the dynamics of critical / noncritical window of EM fracture induced activity is similar with that of healthy / injury heart's electrical action, correspondingly. Notice that we have shown that the noncritical window in the fracture induced EM activity is emerged after the completeness of the "symmetry breaking" phenomenon (Contoyiannis et al., 2005), as it is predicted by criticality. This phenomenon refers to the transition from a state where the whole network is activated, characterized by high ramification, to a state where the activity is restricted to a spatially confined area. There is no direct temporal sequence between the healthy ("p121") and the pathological ("p023") ECGs, since they refer to two different subjects. However, the notion that the symmetry-breaking-type transition of a second order phase transition is a common framework of fracture dynamics and heart's electrical action cannot be excluded.

Herein we draw attention on the delectability of MHz fracture-induced EM emissions. We clarify that all the detected EM emissions are associated with surface EQs which occurred on land or near coastline. We have already clarified this issue (e.g., Kapiris et al., 2003; Eftaxias et al., 2004; Karamanos et al., 2006). We briefly comment on this point focusing on the L'Aquila EQ case.

The L'Aquila EQ was a very shallow land-based event with magnitude 6.3 that occurred on 6 April 2009 (e.g., Eftaxias et al., 2009; Eftaxias et al., 2010). Notice that various methods verify this evidence. Indeed, Walters et al. (2009) using SAR interferometry and body wave seismology constrained the L'Aquila EQ parameters. They found that the top of the activated fault was located from 3.0 km up to 1.5km bellow the Earth's surface. Other methods (see Table 1 in Walters et al., 2009), determine the top of this fault even up to 0 km from surface. We note that this view is further verified by other precursors associated with the L'Aquila EQ. More precisely, satellite TIR anomalies were found 7 days before the EQ (30 March) (Pergola et al., 2010). Boudjada et al. (2010) have found decrease of VLF transmitter signal and Chorus-whistler waves about a week before L'Aquila earthquake occurrence. A ULF (LAI- coupling) EM anomaly was detected from 29 March 2009 up to 3 April 2009 (Eftaxias et al., 2010). Note that the MHz EM anomalies were recorded on 26 March 2009, 2 April 2009 (Eftaxias et al., 2010), and almost 3 days before the EQ event (the presently analyzed anomaly). All the above experimental evidences conduce to the fact that the fracture process was surficial. Therefore, there is no reason why the fracture-induced EM emissions should not be directly launched to the atmosphere without absorption.

Now we focus on the intensity of the emitted EM emission. As it was mentioned, the precursory MHz EM signals are produced during the fracture of the highly heterogeneous media surrounding the fault. For an EQ of magnitude ~6 the corresponding fracture process extends to a radius of ~120km (Bowman et al., 1998). We note that such a large extensive process is necessary in order to observe the other, already mentioned, precursors reported within or around the same time period of the MHz signals emergence. Moreover, as it was mentioned, during the "critical window" the underling fracture process exhibits the critical characterizing the *critical point* of a second order transition, namely, the subunits (cracking area) are well correlated even at arbitrarily large separation, while, appear self-similar structures both in time and space (Stanley, 1987, 1999). The fact that cracking events (EM emitters) are occurred at a large region and correlated strongly enhances the concept that an efficient precursory MHz emission is launched in the atmosphere.

We also draw attention to the fact that a fracture / faulting process is characterized by the formation of fractal pattern. Especially, as it was said, during the critical window appear self-similar structures both in time and space. Fractals are highly convoluted, irregular shapes. The path length between two points is increased by requiring that the small line elements are no longer collinear. This suggests that the number of breaking bonds in a fracture network is dramatically higher in fractal fracture process in comparison to those of Euclidean fracture process. This strongly enhances the possibility to capture these pre-seismic radiations; a network of cracking – electromagnetic emitters having a fractal distribution in space is formed during the fracture of heterogeneous medium. The idea is that an extended emitting efficient Fractal EM Geo-Antenna (Eftaxias et al., 2004) should be formed as a significant surface EQ approaches, launching EM radiation in the atmosphere. A fractal tortuous structure significantly increases the radiated power density, as it increases the effective dipole moment. Recently, the research area known as "fractal electrodynamics" has been established. This term was first suggested by Jaggard to identify the newly emerging branch of research which combines fractal geometry with Maxwell's theory of electrodynamics. The above mentioned idea has been successfully tested (Eftaxias et al., 2004). Finally we note that both the efficiently transmitting part of the Earth's surface and the receiving antennae of experimental station are located on mountainous areas with direct sight link.

## 4. Normal vs. pathological subjects by means of Hurst exponent and wavelet decomposition.

In this section we further study the existence of similarities in the breakdown that accompanies the transition from the critical (healthy) epoch to the noncritical (pathological) one in both biological and geophysical signals under study. The analysis is attempted by means of Hurst exponent and wavelet decomposition which seem to shed more light in the physical background of the observed similar breakdown in the biological and geophysical systems under study.

### 4.1 Focus on the human heart dynamics.

Physiological signals in healthy state present a temporal fractal structure (Ivanov et al., 1999). The present analysis verifies the aforementioned suggestion. This fact indicates that the ECG signal is rooted in fractal forms which consist of self-similar subunits (and sub-sub-units, etc.) that resemble the structure of the overall object. Fractal geometry appears to govern important features of cardiac mechanical function (Goldberger et al., 2002). The self-similar structures support at least one fundamental physiologic function: the rapid and efficient transport over a complex, spatially distributed network.

If a time series is a temporal fractal, then a power-law of the form

$$S(f) \propto f^{-\beta} \qquad (7)$$

is obeyed, where $S(f)$ is the power spectral density and $f$ is the frequency. The spectral scaling exponent $\beta$ is a measure of the strength of time correlations. The healthy ECG under study follows the power law $S(f) \propto f^{-\beta}$. We have proven (Contoyiannis et al., 2005) that the following relation exists between the $\beta$ and $p_2$ exponents:

$$\beta = 5 - 3p_2 \qquad (8)$$

We note that the upper bound for the spectral scaling exponent $\beta$ is $\beta = 2$ by means of the "second-order phase transition" analogy. For the healthy heart the $p_2 -$ exponent is restricted in the region 1.1-1.3 (see Fig. 1c). Based on the relation (8) we conclude that the $\beta -$ exponent values lie in the region 1.1-1.7.

The resulted color-type behavior of the power spectrum density ($\beta > 0$) means that the spectrum presents more power at low frequencies than at high frequencies; the system selects to transmit more power at lower frequencies. If all frequencies were equally important, namely white noise, it would be $\beta = 0$. Moreover, the included electric events are governed by long-range temporal correlations, namely, memory: the current value of the ECG signal is correlated not only with its most recent values but also with its long-term history in a scale-invariant, fractal way. The above mentioned features imply that the data are rooted in mechanism which is compatible with the thermal critical point.

The exponent $\beta$ is related to another exponent, the Hurst exponent $H$, by the formula

$$\beta = 2H + 1 \qquad (9)$$

for the fBm model. $H -$ values can lie anywhere in the range $[0,1]$ ($1 < \beta < 3$) (Heneghan and McDarby, 2000) and characterizes the persistent and anti-persistent properties of the signal according to the following scheme. The range $0.5 < H < 1$ suggests persistence behavior of the signal (super-diffusion), i.e., if the amplitude of fluctuations increases in a time interval, it is expected to continue increasing in the interval immediately following. The range $0 < H < 0.5$ reveals anti-persistency;

i.e., if the fluctuations increase in a period, they probably will continue decreasing in the interval immediately following and vice versa. Values of $H = 0.5$ or $\beta = 2$ indicates no correlation between successive increments. Observation of the particular values implies a transition from anti-persistent to persistent behavior during the evolution of the underlying process.

Based on the Eq. (9), we conclude that the relaxation intervals of the healthy heart are characterized by $H-$exponents distributed in the range $[0.05, 0.15]$. We note that Ivanon et al. (1999) in a pioneering work have studied the human heartbeat dynamics by means of multifractality. They found that for healthy subjects the local Hurst exponents are distributed in the range $[0.07, 0.17]$ in consistency with our results. *The resulted $H-$exponents indicate that the fluctuations in the healthy heart dynamics exhibit strong anti-persistent behavior.* The associated physical information is that the control mechanism regulating the healthy heart operation is characterized by a negative feedback mechanism that kicks the system out of extreme situations. Such a mechanism provides *adaptability*, the ability to respond to various stresses and stimuli of everyday challenges.

The Hurst exponent also describes the roughness of the profile of the time series. The estimated $H-$values $[0.05, 0.15]$ indicate a profile with very high roughness. In the frame of the fBm-model the fractal dimension is given by the relation (Lowen and Teich, 1995; Heneghan and McDarby, 2000):

$$D = 2 - H ; \qquad (10)$$

therefore, the fractal dimension of the time series is close to 2. The $H-$ and $D-$ values show that the fractal structure of the network is characterized by very high ramification which permits rapid and efficiency transport of electric pulses over the complex, spatially distributed in the cardiac tissue network.

The above mentioned relation (8) between the $\beta$ and $p_2$ exponents is valid only in the critical point. Therefore, we determine the Hurst exponent of the pathological ECG in terms of the $R/S$ analysis.

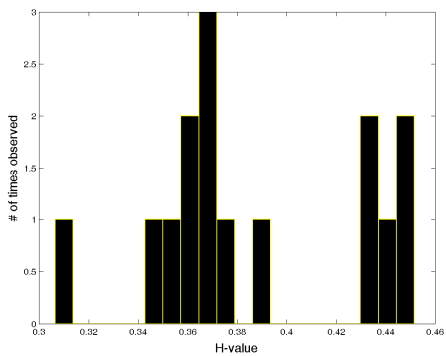
(a)

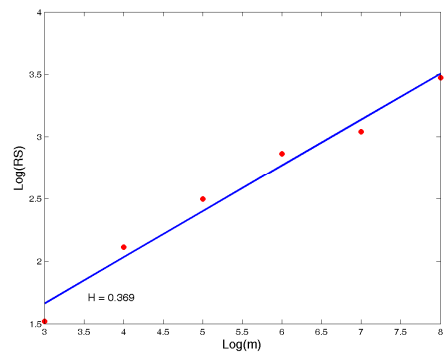
(b)

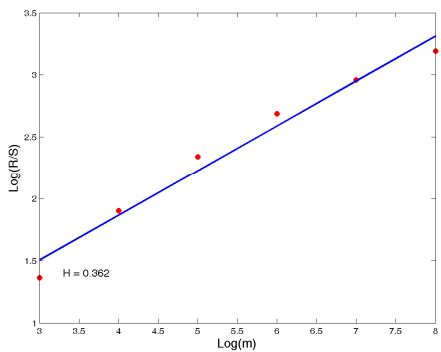
(c)

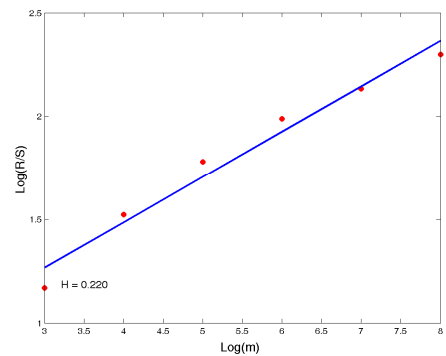
(d)

**Fig. 4.** (a) Histogram of the Hurst exponents resulting for the 15 relaxation parts of the "p023" ECG time-series. (b) Sample application of the $R/S$ method to an indicative relaxation interval of the pathological heart taken from the ECG time-series of the subject "p023" (c) $R/S$ analysis of the total length of the noncritical MHz EM emission excerpt of Fig. 2e associated with the Kozani-Grevena EQ. (d) $R/S$ analysis of the total length of the noncritical MHz EM emission excerpt of Fig. 3e associated with the L'Aquila EQ.

Figs 4a and 4b show that the pathological heart is characterized by $H = 0.387 \pm 0.044$. The control mechanisms regulating the heart dynamics keeps its anti-persistent character, however, they are less anti-persistent than in the healthy dynamics, as the scaling exponent is closer to $H = 0.5$. Its ability to kick the system far from of extreme has been reduced. This result is in consistency with that obtained by Ivanov et al. [1999]. The profile of pathological ECG is smoother in comparison with the healthy one as it is described by a smaller / larger $H$ exponent / fractal dimension. This indicates that the network shows fewer ramifications in its structure. The capability of rapid and efficiency transport over the complex, spatially distributed network has been reduced.

**4.2 Focus on the fracture dynamics.**

As it has been already indicated, if a time series is a temporal fractal, then a power-law of the form of (7) is obeyed, while the Eq. (8) describes the relation between the $\beta$ and $p_2$ exponents. The spectral scaling exponent $\beta$ is a measure of the strength of time correlations.

In the critical window, which is emerged during the damage of the heterogeneous material, the laminar lengths follow a power law distribution with exponent $p_2$ distributed in the range $[1.05, 1.21]$ for the Kozani-Grevena pre-seismic EM emissions (see Fig. 2c) and $[1.04, 1.30]$ for the L'Aquila pre-seismic EM emissions (see Fig. 3c). The power spectral density follows the power law of Eq. (7) where the exponent $\beta-$ is distributed in the ranges $[1.37, 1.85]$ and $[1.10, 1.88]$, respectively, as they are calculated using Eq. (8). The associated $H$ exponent takes values from 0.19 up to 0.43, for Kozani-Grevena, and from 0.05 up to 0.44, for L'Aquila, through Eq. (9), while fractal dimension $D$ takes values in the $[1.57, 1.81]$ and $[1.56, 1.95]$, respectively, as they are calculated using Eq. (10).

The above mentioned results show that the observed critical MHz EM field is originated during the fracture of fractal forms. Elementary self-similar ruptures interact due to overlapping of their (scalar) stress fields. We focus on this point. Fractals are highly convoluted, irregular shapes. The path length between two points is increased by requiring that the small line elements are no longer collinear. This suggests that the number of breaking bonds is dramatically higher in fractal fracture process in comparison to those of Euclidean fracture process. The associated low value of $H$ exponent (close to zero) indicates high roughness of MHz EM time series profile, while the high value of fractal dimension $D$ (close to 2) mirrors that the fracture network is characterized by high ramification. Consequently, a high amount of energy is consumed in risk-free fractures during the damage evolution of heterogeneous material that surround the backbone of strength entities distributed along the fault sustaining the system. This information implies that the nature plays a meaningful $1/f$ music during the earthquake preparation process! On the other hand, the underlying mechanism is characterized by a strong negative feedback mechanism. The above mentioned imply that the critical window includes the crucial feature of "healthy" crust function, namely, adaptability, the ability to respond to external stresses.

Figs 2e – 2h and 3e – 3h show that a dramatic breakdown of critical characteristics, in particular long-range correlations, is appeared during the transition from the critical window to that of noncritical directional fracture; the correlation function turns into a rapid exponential decay. The fluctuations are significantly less anti-persistent than in the critical dynamics, as the scaling exponent is closer to $H = 0.5$. An analysis in consecutive 1024 samples long time windows leads to the result that $H = 0.351 \pm 0.044$ for the Kozani-Grevena case, see also Fig. 4c, and $H = 0.218 \pm 0.028$ for the Methoni case, see also Fig. 4d. The ability of the underlying mechanism to kick the system far from of extreme has been significantly reduced. The profile of the "pathological" MHz EM time series is smoother in comparison with the critical one as it is described by a smaller / larger H exponent / fractal dimension, correspondingly. This indicates that the network shows fewer ramifications in its structure. The capability of rapid and efficiency transport over the complex, spatially distributed network has been reduced.

The above presented analysis shows that significant alterations of fractal dynamics during the transition of fracture process from the "healthy" critical state to the "pathological" noncritical one are similar to those which characterize the alteration of heart dynamics when a healthy heart transitioning to an injury function (Goldberger et al., 2002). A breakdown of fractal physiologic complexity accompanies the transition to the pathological state.

### 4.3 Analysis by means of wavelet decomposition.

The wavelet decomposition reveals physical features that shed light in the transition from the critical to noncritical epoch of heart's operation. Figs 5a and 5b show a scalogram color-coded wavelet analysis of ECG time series in representative healthy and pathological relaxation windows, correspondingly. It is noted that the time-series energy extends to a wider scale (or, equivalently, frequency) range, providing a more "detailed" picture, in the case of the critical epoch, i.e., in the healthy case (Fig. 5a) compared to the heart failure case (Fig. 5b). The same result is even clearly exposed through the energy density of the wavelet marginal spectrum. Indeed, Fig. 5e shows clearly that the energy content of the heart failure case is restricted to higher scales (lower frequencies) only, while the energy content of the healthy heart is distributed to a noticeably wider scale range, indicating that the healthy heart is capable of operating in a much more complicated way responding to many different stimuli. Taking into account that the source size is inversely proportional to the emitted frequency, we infer that the healthy heart's electrical sources are of many different sizes in contrast to the pathological case which are expected to be limited to large sizes only. Therefore, the healthy heart system is characterized by high ramification and adaptability, results that are in consistency to the results obtained through the Hurst exponent and fractal dimension analysis.

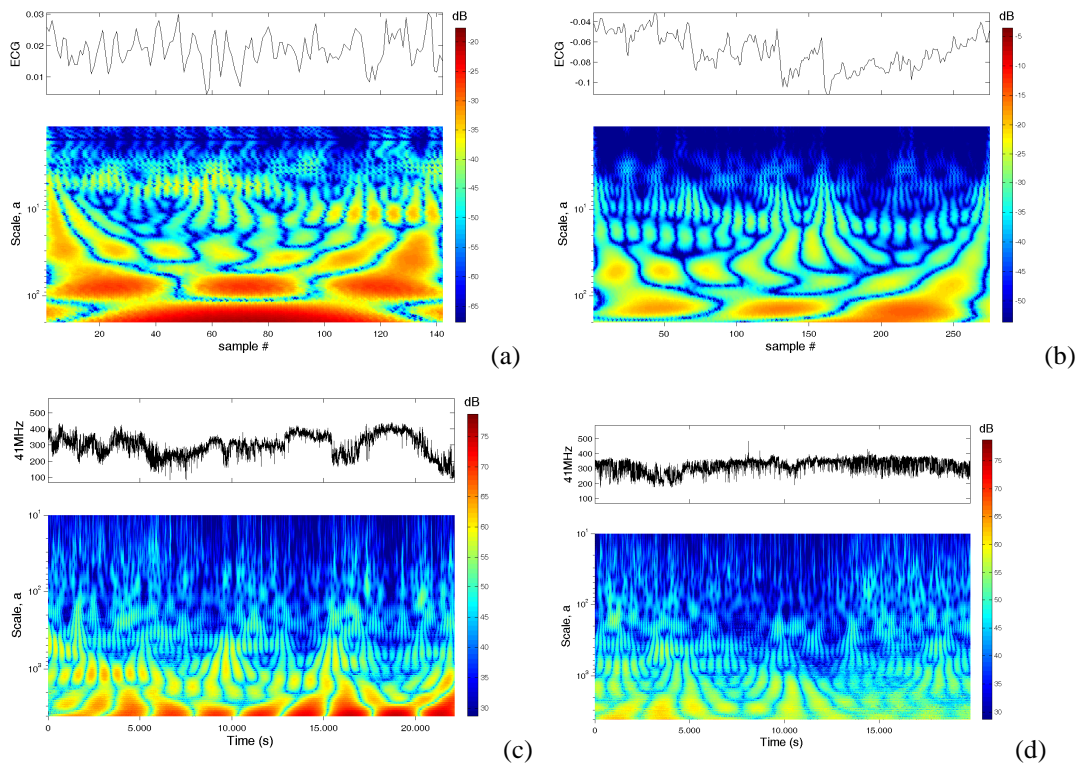

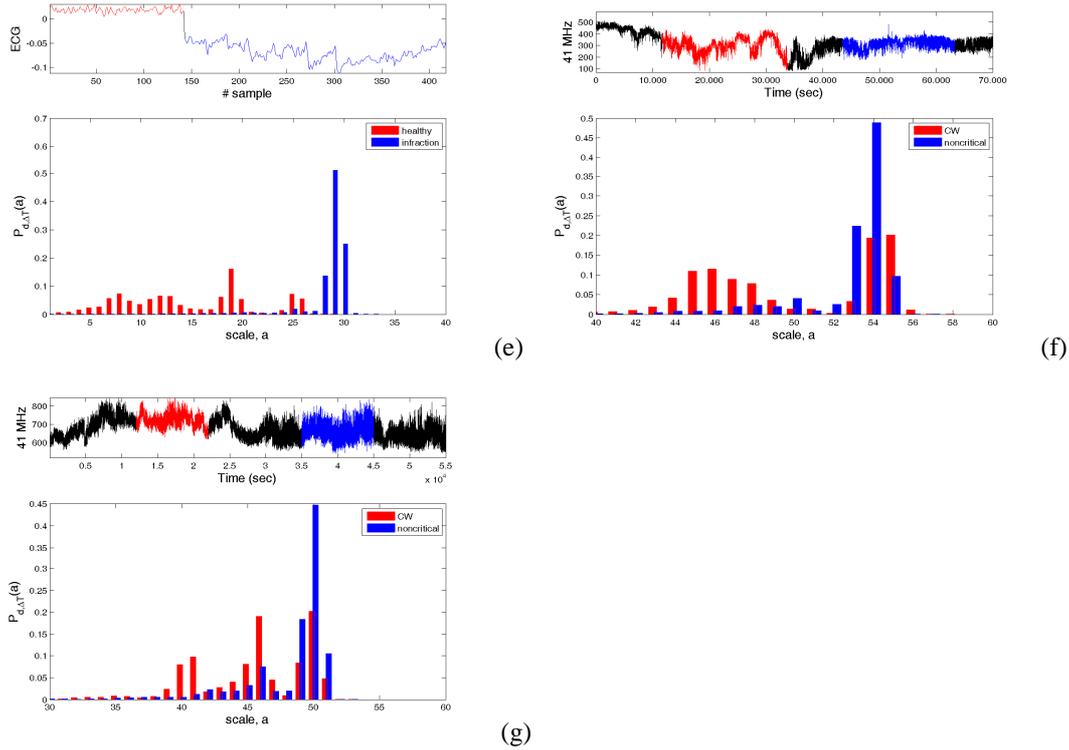

**Fig. 5** (a) Morlet wavelet power scalogram (in dB) of an indicative relaxation interval of the healthy heart taken from the ECG time-series of the subject "p121". (b) Morlet wavelet power scalogram (in dB) of an indicative relaxation interval of the pathological heart taken from the ECG time-series of the subject "p023". (c) Morlet wavelet power scalogram (in dB) of the EM emissions CW associated with the Kozani-Grevena EQ. (d) Morlet wavelet power scalogram (in dB) of the noncritical EM emissions associated with the Kozani-Grevena EQ. Energy density of the wavelet marginal spectrum: (e) of an indicative relaxation interval for each one of the healthy and pathological heart ECG time-series, (f) of the CW and the noncritical excerpt associated with the Kozani-Grevena EQ, and (g) of the CW and the noncritical excerpt associated with the L'Aquila EQ

The wavelet decomposition also enhances the physical picture that emerges during the transition from critical to noncritical state of fracture. Figs 5c and 5d show a color-coded wavelet analysis of MHz EM time series in "healthy" (critical) and "pathological" (noncritical) fracture windows, correspondingly. The critical window (Fig. 5c) and a following noncritical excerpt (Fig. 5d) of the recorded 41 MHz electric field strength (in arbitrary units), associated with the Kozani-Grevena EQ are respectively presented in parallel (time-aligned) to their morlet wavelet power scalograms, with the vertical axis corresponding to the scale, $a$, of the wavelet (time scale, reciprocal to the wavelet "frequency") and the color representing the power spectral level in dB (side colorbar). The common horizontal axis is the time (in s), denoting the relative time position from the beginning of the analyzed part of the EM recording. The brighter colors (towards red, as the colorbars of Figs 5c, 5d indicate) denote larger values of the wavelet power spectral level in dB, corresponding to large fracture fluctuations. Thus, these figures show the evolution of fracture with scale and time. We concentrate on the critical "healthy" window (Fig. 5c). Coherent intense fluctuations are occurred in a wide range of frequencies. The wavelet decomposition reveals a tree like, complex, multi-scale hierarchy to the critical fracture dynamics. However, the system selects to transmit more power at lower frequencies. Higher frequency activities are rooted in small space fracture sources, in contrary, very large sources are recruited during slow fluctuations. These features indicate that the signal is rooted in well ramified fractal forms which are composed of self-similar subunits (and sub-sub-units, etc.) that resemble the structure of the overall object.

In contrast, Fig. 5d shows that the wavelet decomposition on fracture fluctuations from "pathological" fracture reveals the loss of complex multi-scale hierarchy. Coherent intense fluctuations are occurred in a shorter range of frequencies in the noncritical window. The fracture occurs mainly on small space sources. The "healthy" tree-like network of fracture process has lost main large branching patterns. This situation is in consistency with the appearance of directional activity that characterizes a state far from the critical point. The symmetry breaking has been occurred.

The scalogram indicated differences between the critical and noncritical EM emissions, are further elucidated by the corresponding energy densities of the wavelet marginal spectrums as they were calculated for both the Kozani-Grevena and the L'Aquila pre-seismic EM emissions. The results are shown in Fig. 5f and 5g, respectively. Like the heart case, the time-series energy extends to a wider scale (or, equivalently, frequency) range in the case of the critical epoch, i.e., in the "healthy" case compared to the "pathological" (noncritical) case. It is clear that the energy content of the noncritical windows are restricted to higher scales (lower frequencies) only, while the energy content of the CWs are distributed to a noticeably wider scale range, indicating that the "healthy" fracture are characterized by a much more complicated dynamics.

In summary, accumulated evidence enables us to verify the grater complexity of the "healthy" critical dynamics compared to those of "pathological" noncritical conditions into both biological and geophysical systems, as it has already been suggested (Goldberger et al., 2002; Ivanov et al., 1999)

## 5. Conclusions

In the present work we verified the view that a number of complex systems arising from diverse disciplines may have certain quantitative features that are intriguingly similar. The performed analyses reveal the existence of strong analogies between the dynamics of fracture of heterogeneous media at geophysical scale and dynamics of heart. Critical phenomena seem to be the characteristic paradigm to explain the origins of both heart and fracture fluctuations elucidating the mechanisms of their complex dynamics. Characteristics features of a critical point of a second order phase transition were identified in the dynamics of healthy human heart and dynamics of *"critical window"* which appears during the evolution of damage with the time. In both systems, the activated subunits are well correlated even at arbitrarily large separation appearing self-similar structures, fact which is mathematically expressed through power laws expressions for the distributions of spatial or temporal quantities associated with the aforementioned self-similar structures. Especially, the laminar lengths (waiting times) follow a power-law-type distribution. These results are in consistency with previous ones for both systems, i.e., biological (Contoyiannis et al., 2004; Kiyono et al., 2005) and geophysical (Contoyiannis et al., 2005; Contoyiannis and Eftaxias 2009). The fractal dimension $D-$ values of the fractal structure of the networks (electrical conduction / fracture) of both systems show that they are characterized by very high ramification which permits rapid and efficient transport of electric pulses / stresses over the complex spatially distributed in the cardiac tissue / fractured medium, correspondingly. The dynamics of both systems also are characterized by a negative feedback mechanism that keeps the system away from extreme situations. Our analysis showed a common dramatic breakdown of the critical characteristics for the heart-failure subjects or fractured heterogeneous medium in the tail of the fracture process. The alterations refer to: (i) Correlations differences. The distributions of laminar lengths follow an exponential-type law. (ii) The fractal dimensions have been reduced. This indicates that the fractal networks are less ramified. (iii) The electric biological and electromagnetic geophysical fluctuations are less anti-persistent compared to the healthy critical dynamics. The aforementioned found alterations are also in consistency with previous studies for both systems, biological (Ivanov et al., 1999; Kiyono et al., 2005; Varotsos et al., 2007) and geophysical (Contoyiannis et al., 2005; Contoyiannis and Eftaxias, 2009). In summary, our results suggest that dynamics of heart and fracture of heterogeneous media could be analyzed within similar mathematical frameworks.

The Earth is a living planet where many complex systems run perfectly without stopping at all. Activity is constantly taking place not only on the surface, but also on the inside beneath our feet. The earthquake generation is a fundamental sign that the earth is a living planet. The results of the present work show that human-type disease characteristics appear during the earthquake preparation process. The fracture process of the heterogeneous material surrounding the backbone of asperities distributed along the activated fault undergoes a phase transition phenomenon similar to that that undergoes the human heart during its transition from the healthy state to the pathological one. When the fracture of heterogeneous medium is completed the stresses siege the backbone of strong entities distributed along the fault preventing the relative slip of its two faces (Contoyiannis et al., 2004, 2005). The observed kHz EM anomalies signalize the fracture of the aforementioned backbone. In a recent work we have shown by means of nonextensive statistical mechanics the existence of a dynamical analogy between a single fault activation in terms of kHz EM activity / a single seizure activation, namely, a dynamical analogy exists for the ways in which firing neurons / opening cracks organize themselves to produce *a single* epileptic seizure / EQ (Li et al., 2005; Kapiris et al., 2005; Eftaxias et al., 2006, 2012). We have also found that the dynamics of magnetic storms, solar flares, earthquakes and preseismic kHz EM anomalies also show dynamical analogies (Balasis et al., 2011a, 2011b). In summary, the evolution of earthquake preparation process is accompanied by appearance of a heart-failure-type "disease" which is followed by the emergence of an epileptic-seizure-type crisis.

A corollary in the study of complex systems is that by transferring well documented ideas and results from investigations in hitherto disparate areas we can check our proposals and results. This procedure is very useful especially in the study of EM precursors considering the difficulties associated with such factors as their highly complex nature, the rarity of large EQs and subtleties of possible clear EM pre-seismic signatures, the present negative views in literature concerning their existence. The present work extends our proposals concerning the existence of common pathological symptoms in preseismic EM emission and other different well studied "pathological" events further supporting the precursory nature of the observed EM anomalies.

**Acknowledgements**

Research co-funded by the EU (European Social Fund) and national funds, action "Archimedes III—Funding of research groups in T.E.I.", under the Operational Programme "Education and Lifelong Learning 2007-2013".

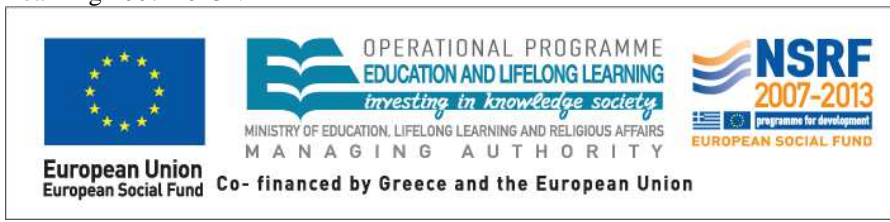